\documentclass[a4paper]{ijisfmt} % For LaTeX2e
\usepackage{times}
\usepackage{graphicx}
\usepackage{doi}
\usepackage[square, comma,numbers, sort&compress ]{natbib}
\usepackage[top=25truemm,bottom=25truemm,left=15truemm,right=15truemm]{geometry}

\title{Psychoacoustic Sonification as User Interface for Human-Machine Interaction}

\author{Tim Ziemer\DAG, Nuttawut Nuchprayoon*, and Holger Schultheis\DAG\DDAG \\\\
\large{\DAG Bremen Spatial Cognition Center, University of Bremen, Germany}\\
\large{\DDAG Institute for Artificial Intelligence, University of Bremen, Germany}\\
\large{* Faculty of Veterinary Science, Mahidol University, Thailand}\\
\large{\{ziemer, schulth\}@uni-bremen.de, nuttawut.nuc@mahidol.edu}
}

\setpapertype{Invited Paper}
\begin{document}

\maketitle

\abst
When operating a machine, the operator needs to know some spatial relations, like the relative location of the target or the nearest obstacle. Often, sensors are used to derive this spatial information, and visual displays are deployed as interfaces to communicate this information to the operator. In this paper, we present \emph{psychoacoustic sonification} as an alternative interface for human-machine interaction. Instead of visualizations, an interactive sound guides the operator to the desired target location, or helps her avoid obstacles in space. By considering psychoacoustics --- i.e., the relationship between the physical and the perceptual attributes of sound --- in the audio signal processing, we can communicate precisely and unambiguously interpretable direction and distance cues along three orthogonal axes to a user. We present exemplary use cases from various application areas where users can benefit from psychoacoustic sonification.


\section{INTRODUCTION}
\begin{figure}[!ht]
%\begin{center}
%\centering
%\setlength{\unitlength}{2.6mm}
\includegraphics[width=0.45\textwidth]{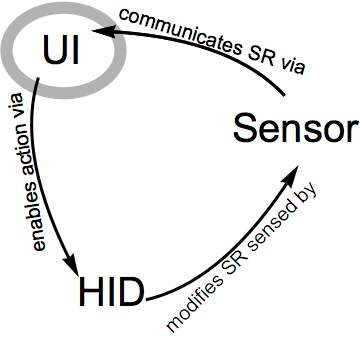}
\caption{Cycle of HMI: Sensors sense information about Spatial Relations (SR), communicate them via User Interface (UI), which enables the operator to take appropriate action via Human Interface Device (HID), which again modifies the SR as sensed by the sensors. This paper deals with psychoacoustic sonification as a UI.}
\label{pic:cycle}
%\end{center}
\end{figure}

Human-machine interaction (HMI) or human-machine collaboration is defined as coordinated interaction of a human and a machine that requires complex sensor-motor control abilities \cite{hcihmi}. Simply put, HMI is a cycle with three stations as illustrated in Fig. \ref{pic:cycle}. A human operator manually controls a machine via a Human Interface Device (HID), which modifies the location and/or orientation of the machine, and thereby its spatial relation to targets or obstacles in the environment. This modification is sensed by the machine's sensors. The sensor data is communicated to the operator via a User Interface (UI). Traditionally, UIs are visual displays. However, in this paper we introduce \emph{psychoacoustic sonification} as an auditory UI for HMI.

Compared to conventional forms of sonification, the strength of psychoacoustic sonification is that multivariate or multidimensional data can be communicated to a user interactively via sound in an unambiguous way with high precision, low technical demands and low latency. This makes it suitable for situations in which vision is limited, or in which the visual scene is overloaded. Throughout this paper we will highlight the benefits of psychoacoustic sonification and present exemplary use cases in which these benefits become effective.

\section{HUMAN-MACHINE INTERACTION}
As illustrated in Fig. \ref{pic:cycle}, Human-Machine Interaction (HMI) \cite{hcihmi} is a cycle with three stations. Human interface devices (HIDs) include computer mouse and keyboard, gamepads, touchpads and touchscreens, knobs and sliders, joysticks, eye-tracking, voice control and gesture control via visual object recognition, motion tracking with optical markers or electromagnetic fields. These serve to operate a machine, vehicle, vessel, spacecraft, aircraft, missile or robot in real, virtual or mixed reality environments.

The machine is equipped with sensors that sense spatial relations between the machine and some objects of the outside world. Sensors for the outside world include radar, lidar, active and passive sonar, electromagnetic sensors, passive or active monoscopic or stereoscopic (infrared) cameras and, e.g., computer vision or motion capture systems \cite{navisens}. Sensors for the location and orientation of the machine itself include tachometer, inclinometers, Hall sensors, accelerometers, gyroscopes, global navigation satellite system (GNSS), compass, or combined Inertial Navigation System (INS) \cite{locationsensor}. Sensor data may be processed and interpreted by a computer processor and communicated to the operator via a User Interface (UI).

Typical UIs are visual displays \cite{vdisplay}, like  lamps, seven segment displays,  %liquid-crystal displays (LCD), light-emitting diode (LED), 
and monitors that show camera recordings, graphs, maps, charts, depictions, icons, text, crosshairs, meter needles, and/or tables, etc. %Haptic displays \cite{haptic} include refreshable braille displays and devices that create vibrations, resistance, and/or force-feedback. 
Auditory displays \cite{davidbuch} are less common UIs \cite{hciindi,hcihb}. They represent data, e.g., by means of speech, auditory icons, earcons, and the core element: sonification.

\section{AUDITORY DISPLAYS}
\emph{Auditory Displays} (ADs), sometimes referred to as \emph{tactical audio}, are sounds that inform a user during HMI. In that sense they serve as UIs that can communicate sensor data to the operator. An overview can be found in \cite{davidbuch,thomas}. ADs can have multiple elements.

\emph{Auditory icons} are sounds that mimic a real-world sound that is causally or symbolically related to the represented data. For example the sound of slamming a door can represent closing a file. Their major advantages are that they can be intuitively understood and that they can be easily recorded or synthesized. The main drawbacks are that they carry only a very limited amount of discrete information, i.e., nominal scale of measure.

In contrast to auditory icons, \emph{earcons} are the combinations of discrete sounds to short sound events that are not derived from the real world. For example some computer operating systems play a certain melody when attaching a device via USB, and a similar melody with the same instrument when detaching a device. The major advantage of earcons is that they can carry more information than auditory icons, like hierarchies, i.e., ordinal scale of measure. Their drawbacks are that their meaning needs to be learned, no continuous information can be presented and overlapping earcons may be indiscriminable.

\emph{Speech} is self-explaining. Spoken words or sentences inform the user about the data. Famous examples of speech as a UI are screen readers. The major benefit of speech is that even complicated information and relations can be communicated. The most crucial drawbacks of speech are that it takes much time, and that it is cognitively demanding to interpret language. Only a small number of data points per second can be expressed by words, no continuous data streams.

In contrast to earcons, \emph{sonification} is not discrete in nature. It can therefore communicate continuous data. A famous example of sonification is auditory pulse oximetry, where the oxygen concentration in the patient's blood is communicated to the anesthesiologist by means of the pitch of a complex tone that is triggered every heartbeat. The main advantage of sonification is that it can potentially present the highest density of information. It can carry much more information than auditory icons and earcons, much faster than speech. Sonifications can represent data on an interval or even ratio scale of measure. The main disadvantage is that the meaning of each individual sonification needs to be learned by the user. Furthermore, sonifying a high amount of data often lead to ambiguous or confusing results.

Auditory displays are typically employed in situations in which visual displays are (partly) occluded or lie outside the visual focus or completely out of the visual field  \cite{hciindi,hcihb}. Furthermore, they are useful when vision is limited, e.g., due to environmental conditions, like darkness, fog, smoke, muddy waters, sandstorm. They also complement visual displays to overcome  shortcomings of visual UIs, like the missing depth on monitors, a visual overload due to too many depictions, or  missing detail, e.g., due to graphics overlay. Auditory displays can be useful alternatives to graphical displays for visually impaired or blind users. Auditory displays are also useful in situations in which visual perception is overloaded, e.g., by the environment or by a high number of visual displays, or when the visual channel is needed for tasks other than HMI, like exploring the environment. Last but not least, reaction times to sound are typically much shorter compared to visual cues \cite{morris} which is why audible alarms \cite{alarmreview} are ubiquitous. Not only can auditory displays serve as an alternative to visual displays, they can also complement or augment visualization in terms of a multi-modal user interface.

\section{PSYCHOACOUSTIC SONIFICATION}
Sonification is the communication of data by means of sound. The translation from input data to output sound is called \emph{mapping}. In contrast to auditory icons, earcons, and speech, sonification is able to communicate one continuous data stream. Hence, it helps for data monitoring of stock markets and vital functions, and for navigation aid in driving, piloting and surgery \cite{davidbuch,jmui}.

For tasks that require complex information, i.e., multivariate or multidimensional data, conventional auditory displays \cite{123channels2,matti,Black2017,komatsu} could be demonstrated to be helpful in supplementing visual UIs. However, when trying to sonify multiple data streams, no conventional auditory display  can satisfy all demands, which are

\begin{enumerate}
\item interpretability
\item linearity
\item continuity
\item high resolution
\item absolute magnitudes (interval or ratio scale of measure)
\end{enumerate}
of each single data stream and
\begin{enumerate}
\item orthogonality 
\item integrability%combinability 
\end{enumerate}
of multiple data streams.

\begin{figure}[!ht]
%\begin{center}
%\centering
%\setlength{\unitlength}{2.6mm}
\includegraphics[width=0.45\textwidth]{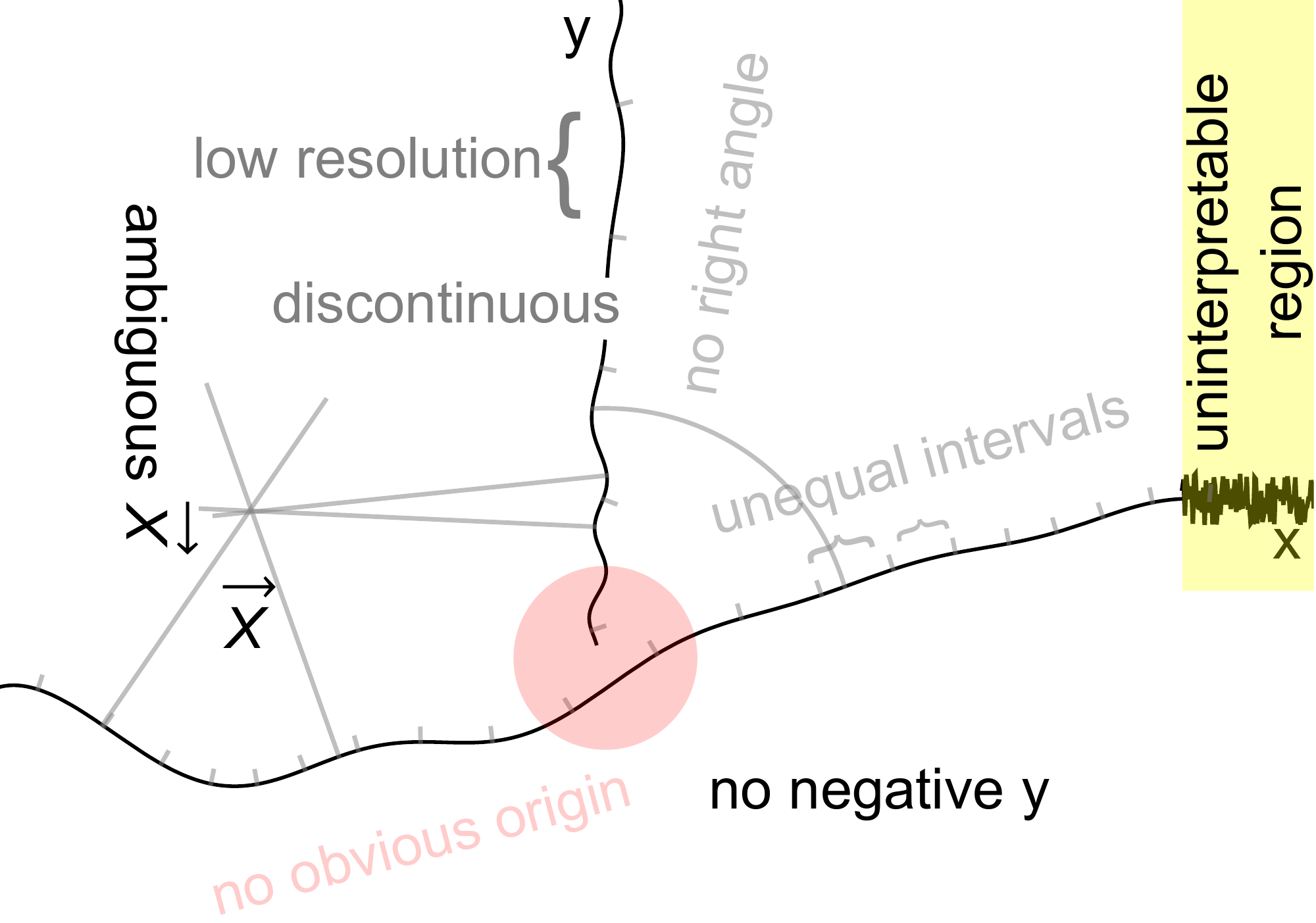}
\caption{Typical perception issues in  conventional, multidimensional sonification.}
\label{pic:weak}
%\end{center}
\end{figure}

Figure \ref{pic:weak} illustrates common issues in multidimensional sonification. First of all, the $x$-axis is partly chaotic, which makes the sonification \textbf{uninterpretable} in that region. At the same time, the $y$-axis is \textbf{discontinuous}. This creates audible artifacts instead of a smooth transition when raising $y$ continuously. Furthermore, the $y$-axis has \textbf{no negative} part. This means, it is at best half a dimension, since it only represents a distance in one direction, not in two. The low number of ticks on the $y$ axis represents a \textbf{low resolution}, which is a typical issue in sonification. The two axes are no lines but curves. This curvature is a metaphor for the circumstance that the axes are nonlinear. This nonlinearity implies \textbf{unequal intervals}, i.e., doubling $x$ does not necessarily sound twice as intense. The two axes do not cross, i.e., the coordinate system has \textbf{no obvious origin}. The two axes are \textbf{not right-angled}, i.e., the axes are not orthogonal. This means the axes can at most be considered as fractal dimensions. Due to the curvature of the axes, and their lack of orthogonality, many points $\vec{X}$ in the coordinate system are not unique, i.e., 
they are \textbf{ambiguous}. This means, when moving the vector $\vec{X}$, it is impossible to recognize whether the $x$- value, the $y$-value, or both has/have changed.

Psychoacoustic sonification \cite{asa,cars,dgm} is the first sonification technology that satisfies all of the above-listed demands. Psychoacoustics describes the relationships between the physical, acoustic quantities and the perceptual, auditory qualities of sound by means of nonlinear equations \cite{psy,fire}. Physical quantities include amplitude, frequency and phase over time and space, auditory qualities include pitch, beats, roughness, sharpness, fullness and tonalness. These qualities have been shown to be able to accurately capture human listening listening experience.%Pitch is considered multi-dimensional, consisting of rectilinear height and cyclic chroma. 

Practically all physical sound parameters can affect each auditory quality. Consequently, physically quantities are not perceptually orthogonal, i.e., independent auditory qualities. In contrast to conventional sonification, psychoacoustic sonification does not map input data to physical sound quantities, but to perceptual auditory qualities. Psychoacoustic considerations in the signal processing improve the interpretability, linearity and resolution of sonification and make orthogonal and integrable multivariate or multidimensional sonification possible.

\begin{figure}[!ht]
%\begin{center}
%\centering
%\setlength{\unitlength}{2.6mm}
\includegraphics[width=0.45\textwidth]{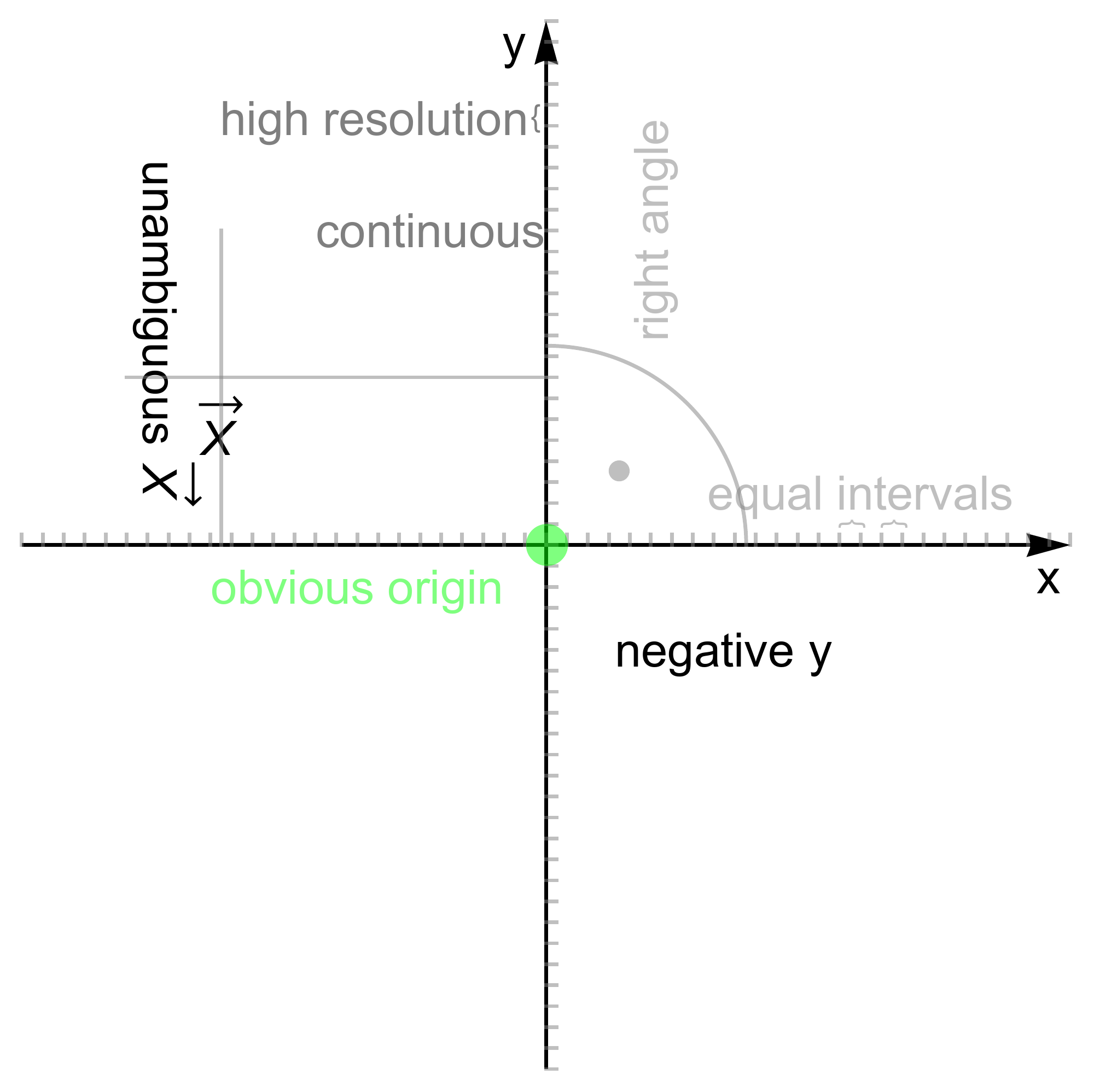}
\caption{Perception of psychoacoustic sonification.}
\label{pic:strong}
%\end{center}
\end{figure}

Figure \ref{pic:strong} illustrates psychoacoustic sonification. Here, the sound impression is clear and non of the above-mentioned issues occur.

\begin{figure}[thbp]
%\centering
\includegraphics[width=0.45\textwidth]{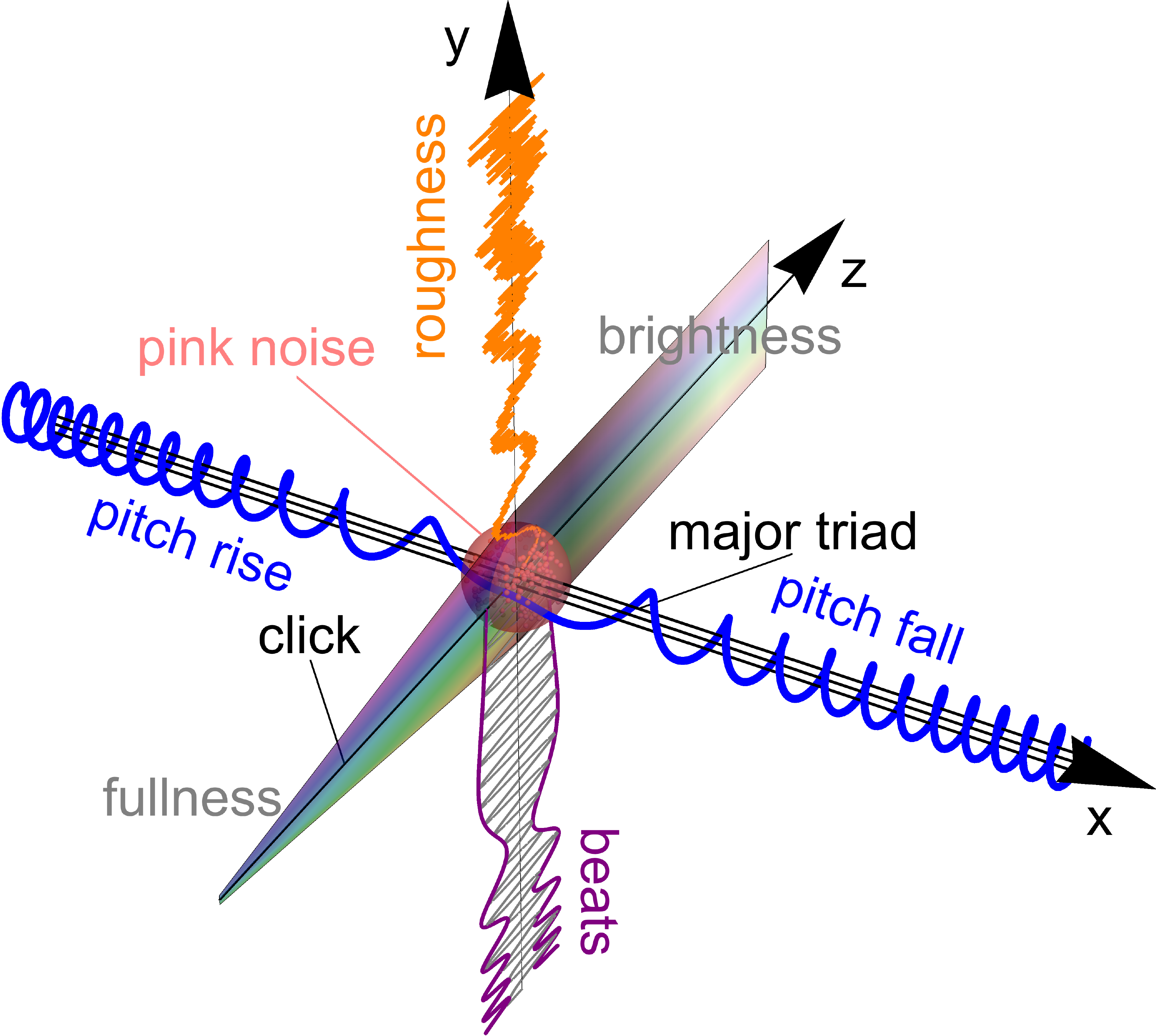}
\caption{This psychoacoustic sonification uses cyclic pitch change as $x$-dimension, speed of beats and degree of roughness as $y$-dimension and degree of sharpness and fullness as $z$-dimension. A click indicates surpassing the $x$-$y$ plane, a major triad surpassing the $x$-$z$ plane, pink noise indicates that the target region has been reached.}
\label{pic:graphics}
\end{figure}

An example of a psychoacoustic sonification is illustrated in Fig.~\ref{pic:graphics} for three-dimensional data. The sonification is a monophonic, tonal sound. Its \emph{attributes} or \emph{characteristics} represent different directions. The \emph{magnitude} or \emph{intensity} of each attribute represents the distance along that direction. The target lies at the origin of the coordinate system at $\vec{X}_\mathrm{target}=(0,0,0)$, a vector $\vec{X}_\mathrm{user}$ describes the location of the user.

The direction of the target relative to the user along the $x$-axis is mapped to the direction of the chroma aspect of pitch. At $x=0$, pitch is steady, $x>0$ creates a falling pitch, $x<0$, creates a rising pitch. The distance along the direction is mapped to the speed with which pitch rises or falls. The larger the absolute value, the faster the pitch rise or fall.

The $y$ dimension is divided in two. Positive $y$-values are mapped proportionally to the intensity of roughness. Negative $y$-values are mapped to the speed of beats, i.e., regular loudness fluctuations. Only at $y=0$ the sound exhibits neither roughness nor beats.
\\
\fbox{\begin{minipage}{24em}
Some simple trajectories through this two-dimensional sonification space can be found on our YouTube channel at \url{https://tinyurl.com/ycwmdh8r}.
\end{minipage}}

The $z$-dimension is also divided in two. Positive $z$-values are mapped to the degree of brightness, negative values to the degree of fullness. At $y=0$ the sound is dull but full.
\\
\fbox{\begin{minipage}{24em}
Some exemplary trajectories through a three-dimensional psychoacoustic sonification can be found at \url{http://tinyurl.com/y4um5odf}. The example videos slightly deviate from the mapping principle explained above.\end{minipage}}

In a navigation task the origin of the coordinate system represents either the target to reach, or the obstacle to avoid. The sonification itself informs the operator whether the origin has been reached or not. Only at the origin the sonification has a steady pitch and loudness, sounds smooth, dull and full. No reference tone is needed. At all other locations the sound attributes inform the operator where the target or obstacle lies. 
One important characteristic of this sonification is that it is intuitive on a qualitative level. Parameters like brightness and roughness, have an innate valence to a human listener. Near the origin, the sound is comparably pleasant and calm. However, towards the extreme values, when the operator is far away from the target, the sound becomes more and more obtrusive, unpleasant, and urgent, forcing the operator into action. For example, quick chroma fluctuations sound like a siren, a high degree of roughness sounds scratchy and distorted.
\\
\fbox{\begin{minipage}{24em}
The urgency of extreme values in the chroma- and beats/roughness-dimension can be heard in a YouTube demo video: \url{https://youtu.be/Z_Iu575erI4}.\end{minipage}}

In addition to the interactive sonification, earcons are triggered. A click informs about surpassing the target height, a short major triad informs about surpassing the target depth. Reaching an extended zone around the ideal target point triggers subtle pink noise.

Together, the psychoacoustic sonification and the additional sounds of the auditory display interactively inform an operator about the location of a sensed or defined target in three-dimensional space. For example, the location of the machine can be the origin of the coordinate system. Then, the sound attributes inform the operator about the direction along the three spatial dimensions ($x$, $y$ and $z$) and about the distance along each direction. So the operator gets informed whether the target has been reached, and if not, where the target is located.

We have experimentally validated the first two dimensions in multiple passive and interactive studies \cite{cars,asa,dgm,poma,acta,jmui,icad,daga,dagaabstract}. Passive listening tests with the $x$-$y$ dimensions revealed that novice users can interpret the direction and magnitude of the two dimensions unambiguously after just $5$ minutes of training \cite{poma}. Experiments with interactive sonification where the sound is continuously modified by the action of the operator revealed that interaction improves interpretability and precision a lot.

\begin{figure}[thbp]
%\centering
\includegraphics[width=0.45\textwidth]{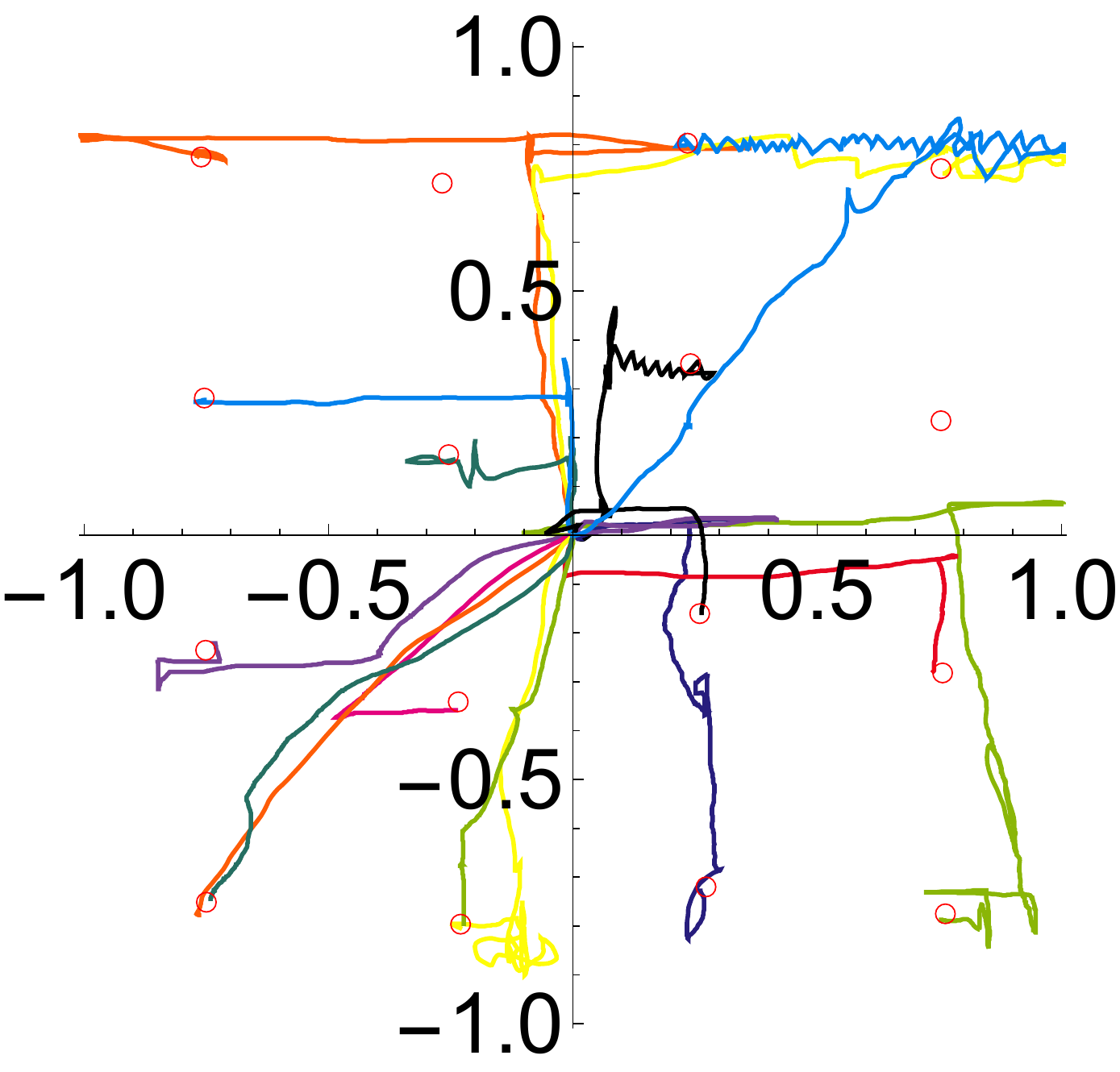}
\caption{Typical mouse trajectories from novice users of psychoacoustic sonification. Some users approach the target axis-by-axis, whereas others aim for the direct path. Insecure users trigger the click repeatedly to confirm that they are still at the target height.}
\label{pic:trajs}
\end{figure}

The signal processing for an interactive psychoacoustic auditory display including sonification and earcons is described in \cite{acta}. Interactive experiments revealed that novice users accurately find invisible sonified targets that have a diameter of just $4$ mm on a two-dimensional computer screen with an area of $4000$ mm${}^2$ \cite{icad,jmui}. They only needed $30$ minutes of explanation and exploration of the sonification to achieve this. Typical mouse trajectories are illustrated in Fig. \ref{pic:trajs}. Practically all targets have been found by each participant. The six trajectories on in the lower right quadrant approach the target almost on the direct path. Direct paths highlight that users were able to integrate the two auditory dimensions and derive the respective angle from the two Cartesian axes. Other participants approached the target axis-by-axis, like the four examples in the lower right ($x$ first) and the three examples in the upper left quadrant ($y$ first). This strategy underlines that the two dimensions are orthogonal, i.e., unambiguously interpretable and distinguishable, even though the two dimensions are presented at the same time. Some participants sometimes surpassed the target height over and over, like the three examples in the upper right quadrant. This action triggers the click repeatedly and confirms that the users are still at the target height.

In \cite{icad2019} we presented the signal processing for the third dimension. Initial results can be found in \cite{iwin}, the complete results have been submitted to \cite{hf}. A preprint is available \cite{preprint}.

\begin{figure}[thbp]
\centering
\includegraphics[width=0.45\textwidth]{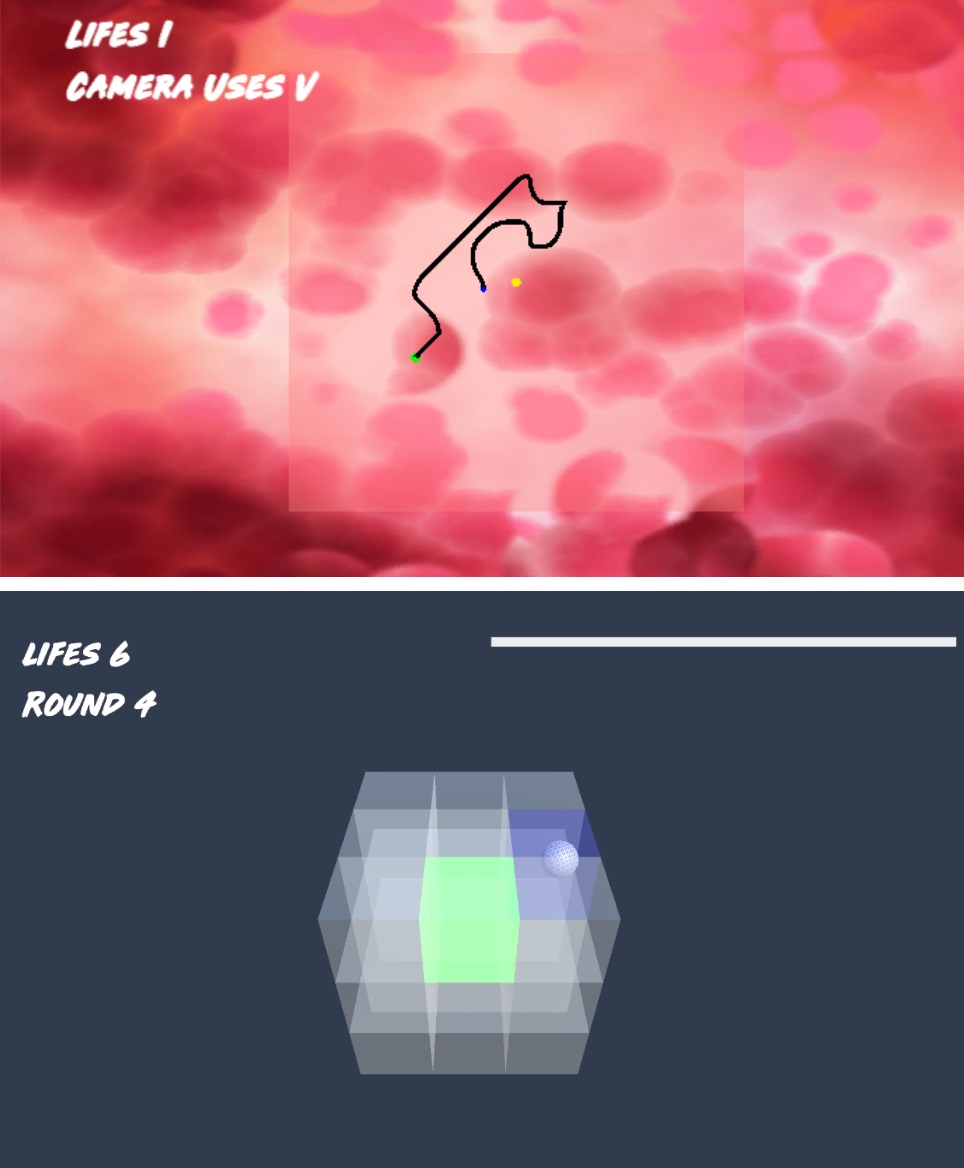}
\caption{CURAT visual game environment to evaluate the 3D psychoacoustic sonification.}
\label{pic:curat}
\end{figure}

In a game-like environment we are validating the 3D psychoacoustic sonification. Screen shots of this serious game are shown in Fig. \ref{pic:curat}. People can play the game for fun, while we collect their motion trajectories to receive statistics from a large population and longitudinal data about learning effects. Game elements, like high scores, levels, missions and power-ups motivate the players to keep learning the sound metaphors, improve their skills and play for a long time. At higher levels less visual cues are given, while more information is presented auditorily to the player. At some point the player depends completely on the sound to navigate through three-dimensional space. Details on the game design and progress are available on \url{http://curat.informatik.uni-bremen.de/}. 

\begin{figure}[thbp]
\centering
\includegraphics[width=0.45\textwidth]{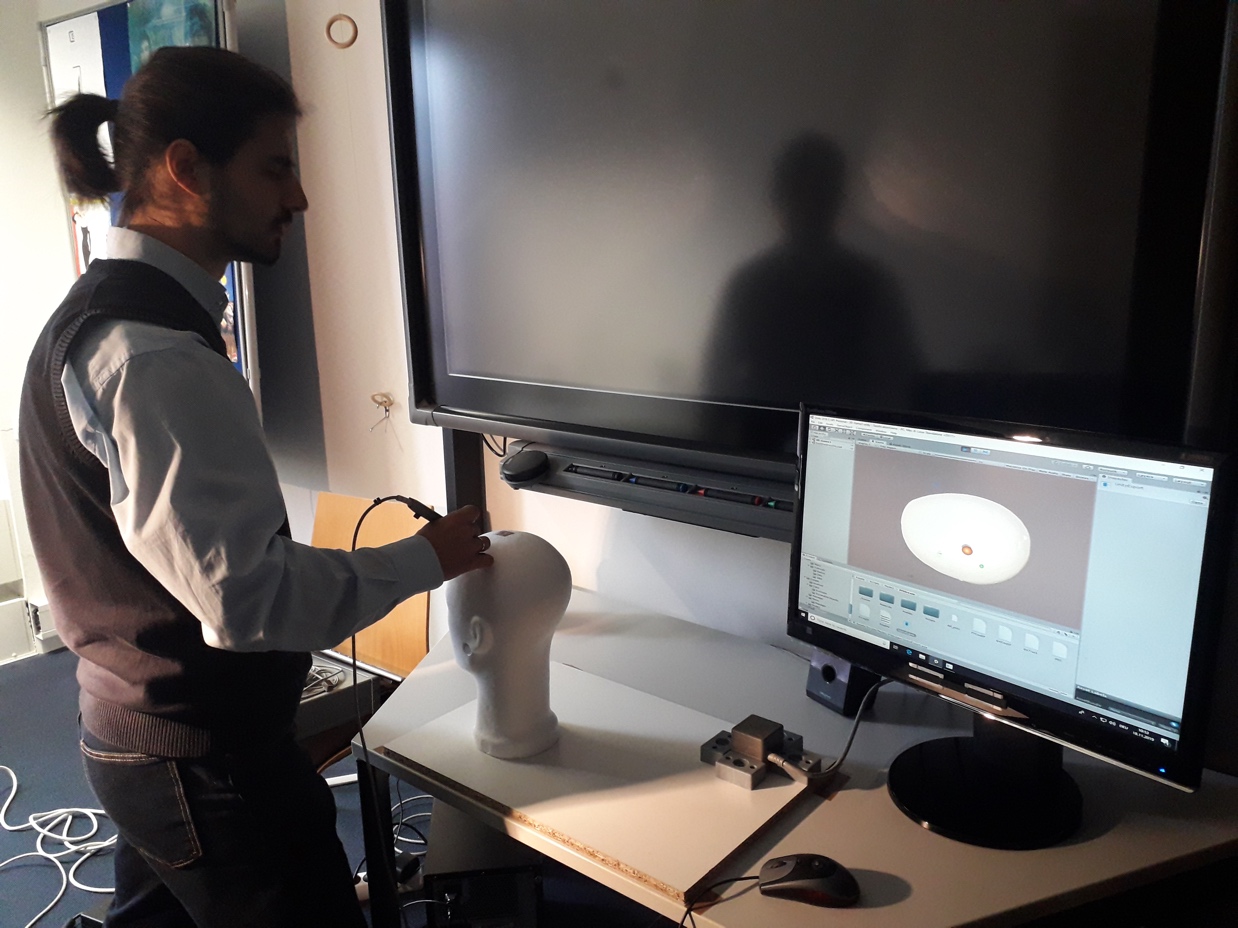}
\caption{Tim Ziemer testing the 3D psychoacoustic sonification in a surgical setting.}
\label{pic:craniotomy}
\end{figure}

Figure \ref{pic:craniotomy} shows how we test the 3D psychoacoustic sonification in a surgical scenario. The planned location of burr holes for a craniotomy are found by means of sound. In a craniotomy, a part of the skull is removed, e.g., to insert an ablation needle that destroys a brain tumor. Here guidance is necessary, because a skull hardly provides visible landmarks that help for orientation. At the same time it is crucial to remove exactly the planned part of the skull to ensure that the needle can be inserted at the planned location, so that the tumor can be reached via the planned insertion angles. However, guidance by means of visual UIs has a number of drawbacks, which are discussed in detail in Section \ref{surgery}. 

So far, we have implemented three dimensions and are in the process of evaluating them. However, psychoacoustic sonification is not limited to three dimensions. The current 3D sonification is completely tonal and largely harmonic. As we have already demonstrated, noisy and impulsive elements, i.e., pink noise and clicks, can be added without degrading the 3D sonification. We believe that at least three additional dimensions are implementable, with proper psychoacoustic sound design of noise and/or pulses. 

%Better mapping, linearity, continuity, (absolute 0, i.e., ratio scale), orthogonality, interpretability, precision/resolution, eyes-free/blind, in darkness, fog, smoke, muddy waters, out of sight, no occlusion
In the next section, we describe application areas and potential use cases where psychoacoustic sonification can act as interface for human-machine interaction.

%\section{SONIFICATION}
%What is psychoacoustic sonification
%psychoacoustics, acoustics
%multidimensional, multivariate, orthogonality, continuous, resolution, 

%\emph{psychoacoustic sonification}, an interactive sound that enables the operator to continuously keep track of multiple pieces of information with high precision. Here, psychoacoustics is of particular importance; it .
%Unmanned aerial vehicle (UAV)%!?!
%multidimensional navigation%!?!

\section{USE CASES}
Psychoacoustic sonification can serve as a UI to communicate the position of a 
point in a coordinate system. The point in the coordinate system may be a location in space or a vector in any other multi-dimensional or multivariate space. The position is either sensed by means of sensors, or it is pre-defined. The point can either be a natural limit, a target point to reach, or a critical point to avoid. The point and the origin of the coordinate system can either be fixed, or dynamic. As mentioned above, the psychoacoustic sonification can serve as a complementary or substitutional UI for HMI in situation in which visual UIs are impractical or insufficient.

As these conditions seem rather abstract, we present some exemplary use cases from different domains that highlight, how machine operators can benefit from psychoacoustic sonification.

\subsection{Drone Flying}
In manual drone flying, the HID is a remote control that has two sticks to control pitch, throttle, yaw and roll, and buttons that trim pitch to move back and forth, throttle to increase or decrease, and yaw or roll to move sideways. Practically all the above-mentioned sensors are available for drones, including GNSS and INS to track self-location and orientation and radar, lidar and sonar to track obstacles in the environment.

\begin{figure}[thbp]
\centering
\includegraphics[width=0.45\textwidth]{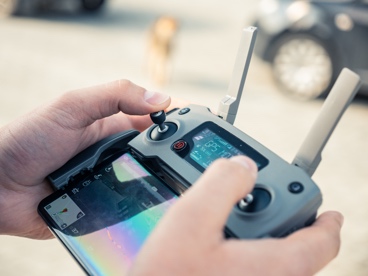}
\caption{Typical drone remote. A visual UI shows the camera view overlain by graphics that indicate spatial relations, like drone coordinate and orientation, and distance and direction of obstacles.}
\label{pic:remote}
\end{figure}

Visual UIs for manual drone navigation are often included in the remote control. They show the camera view overlain by depictions of compass, GPS location on a map and other graphics that indicate the location of obstacles, as illustrated in Fig. \ref{pic:remote}.

\begin{figure}[thbp]
\centering
\includegraphics[width=0.45\textwidth]{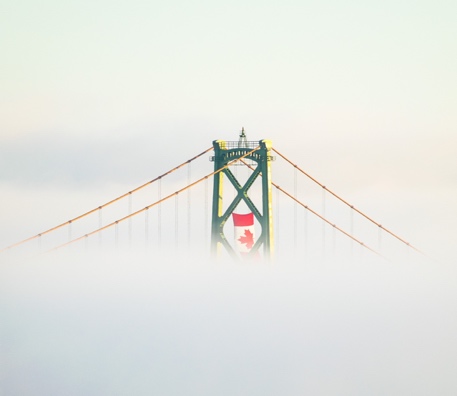}
\caption{In fog, visual navigation is almost impossible. Here, sensors can scan the environment and UIs can inform the operator about the location of obstacles.}
\label{pic:fog}
\end{figure}

Manual Drone flying without sensors is restricted to situations in which the drone is in the visible field of the operator, or in open territories. Uncertainties occur, for example, in fog, as illustrated in Fig. \ref{pic:fog}.

Navigation by the drone's camera is only safe for basic flying, if the drone flies in the direction of the camera view, since obstacles outside the camera view are overseen. Furthermore, transferring the camera view to the remote control introduces a latency. This enforces the operator to slow down and wait for the picture when approaching an obstacle. For advanced maneuvers, some spatial relations need to be sensed. Autonomous drones are largely capable of flying without user control. However, in the case of uncertainty, e.g., due to environmental conditions or tasks that seem unsolvable to the machine's artificial intelligence, human control is necessary, and the operator can fly, while the drone supports her with the available sensor information. In such cases, e.g., in regions with a high density of buildings, trees, transmission lines and other structures, as in urban areas, forests, factory premises, and inside hangars and other buildings, UIs that inform the operator about sensed obstacles help avoiding a crash.

In visual UIs, overlays can occlude important aspects of the camera view. When visualizing the camera view and the location of obstacles on two different screens, the operator has to decide which screen to consult. Here, psychoacoustic sonification may be preferable over graphical UIs in manual drone navigation.

\begin{figure}[thbp]
\centering
\includegraphics[width=0.45\textwidth]{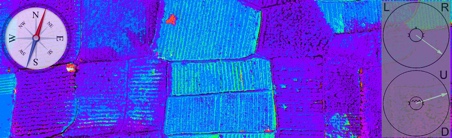}
\caption{Near infrared spectrometry view from a drone in agricultural planing and monitoring. Overlays indicate the location of obstacles, but they partly occlude the view.}
\label{pic:agriculture}
\end{figure}

In agricultural planing and monitoring with drones, near infrared spectrometry is used to identify the healthiest spots (red) and unhealthy crops (yellow). Here, a visual UI is realized with overlays that allow the operator to avoid hitting obstacles, like overhead transmission lines or trees. However, the overlays occlude some of the spectrometry visualization, as illustrated in Fig. \ref{pic:agriculture}. Here, psychoacoustic sonification allows the operator to concentrate on the visualization while being able to avoid sensed obstacles.

\begin{figure}[thbp]
\centering
\includegraphics[width=0.45\textwidth]{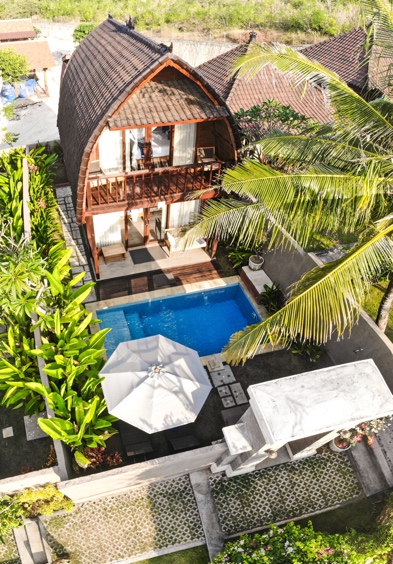}
\caption{Finding the perfect camera view without distracting graphics.}
\label{pic:motif}
\end{figure}

Drones are often used for photography, as illustrated in Fig. \ref{pic:motif}. Here the drone operator visually concentrates on finding the perfect spot for a camera shot. Autonomous drones might prevent the photographer from finding the perfect spot, if it lies dangerously close to an obstacle. However, for manual navigation the camera only helps to avoid obstacles in front. But it is likely that the operator will fly the drone backwards, up, down, left and right, while focusing the camera on the motif. Here, visual UIs with overlays of sensed obstacle positions are disruptive. Two-screen solutions without overlay are dangerous, because the visual focus lies on the camera view and not the navigation aid. In this situation, psychoacoustic sonification allows the operator to fly safely while finding the right spot.

\begin{figure}[thbp]
\centering
\includegraphics[width=0.45\textwidth]{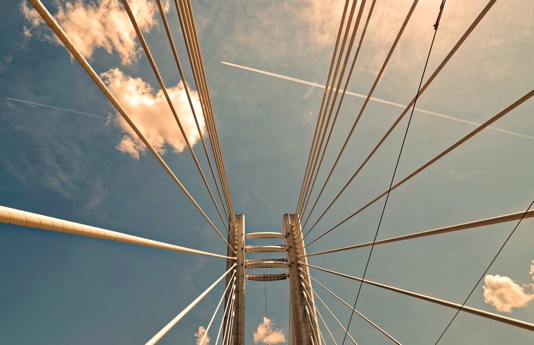}
\caption{The location and orientation of drones constantly changes in inspection tasks. Obstacles can be everywhere, but the visual focus needs to lie on the inspected structure.}
\label{pic:bridgeinspection}
\end{figure}

Visual inspection is a task similar to photography. Visual inspections are carried out, e.g., for airplanes in hangars or roofs in urban areas or factory premises. Figure \ref{pic:bridgeinspection} shows an exemplary drone view of a bridge inspection. Here, obstacles are omnipresent, like abutments, piers, girders, trusses, cables. But manual navigation may become inevitable when the autonomous drone mode cannot find a proper perspective. The visual focus needs to lie on the camera view. Other graphical UI elements are distracting. Again, psychoacoustic sonification allows a single operator to both navigate the drone and perform the visual inspection at the same time.

\begin{figure}[thbp]
\centering
\includegraphics[width=0.45\textwidth]{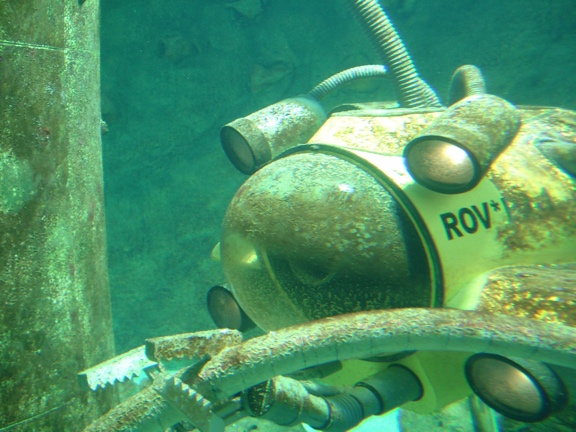}
\caption{Underwater robot for maintenance work. Psychoacoustic sonification can help reaching the target structure and avoiding potentially moving obstacles despite limited vision.}
\label{pic:ur}
\end{figure}

Even though we discussed practical use cases for psychoacoustic sonification in the field of drone navigation, the principle transfers to all kinds of unmanned aerial vehicles, rockets and missiles, and even to other remotely operated vehicles (ROVs), like unmanned surface vessels and unmanned underwater vehicles. Figure \ref{pic:ur} shows an underwater robot.

Underwater robots  \cite{ur} are mostly used for survey work, like installation, inspection, welding and maintenance of subsea structures and production facilities. Especially offshore oil and gas installations are almost exclusively serviced by ROVs \cite{ur}. Here, proximity sensors detect obstacles based on active sonar, while a so-called \emph{short baseline acoustic positioning system} provides absolute location information, just like GNSS, which are not working underwater. Underwater, vision can be limited due to darkness and muddy waters, so redundant multisensor systems are employed to navigate safely to the target structure. Here, psychoacoustic sonification can give a low-latency feedback on the location of obstacles.

\subsection{Manned Aircraft}
\begin{figure}[thbp]
\centering
\includegraphics[width=0.45\textwidth]{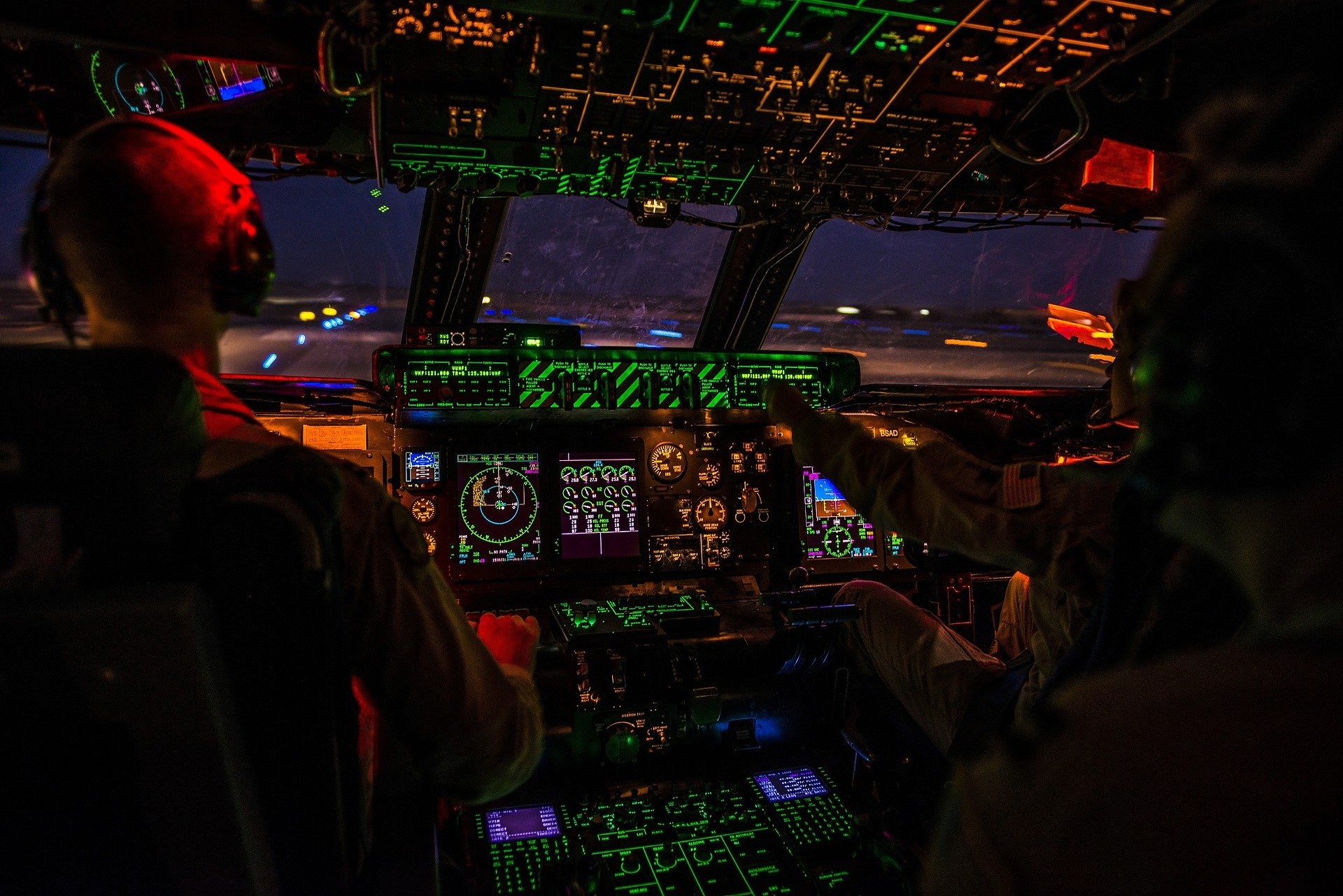}
\caption{Cockpit of an airliner. Not all visual displays lie in the visual focus. Some may even be occluded by the pilot or the copilot.}
\label{pic:civil}
\end{figure}

\begin{figure}[thbp]
\centering
\includegraphics[width=0.45\textwidth]{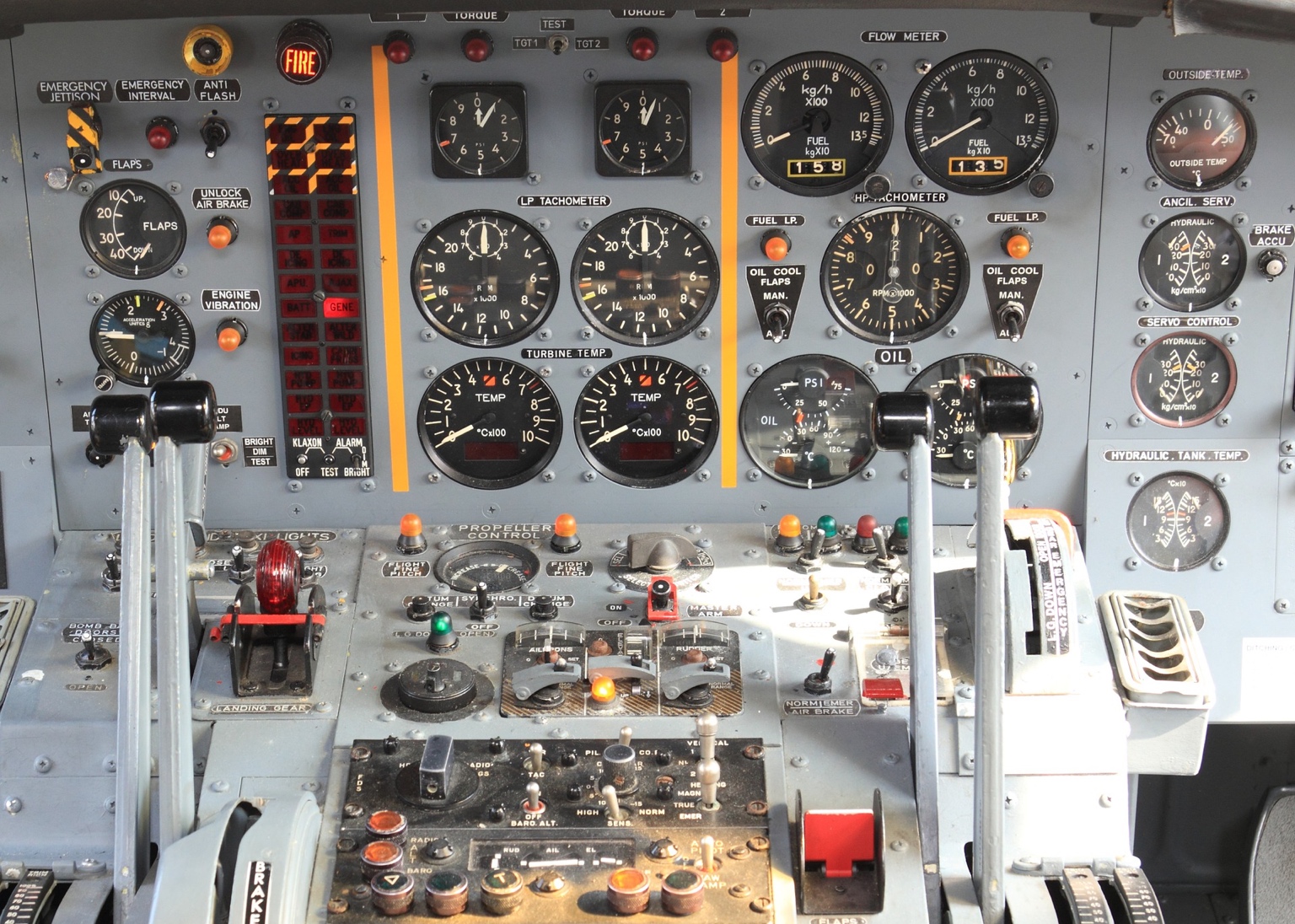}
\caption{Cockpit of a jet fighter. Some maneuvers cause a mismatch of vision and balance and may even impair the ability to read visual displays.}
\label{pic:fightjet}
\end{figure}

\begin{figure}[thbp]
\centering
\includegraphics[width=0.45\textwidth]{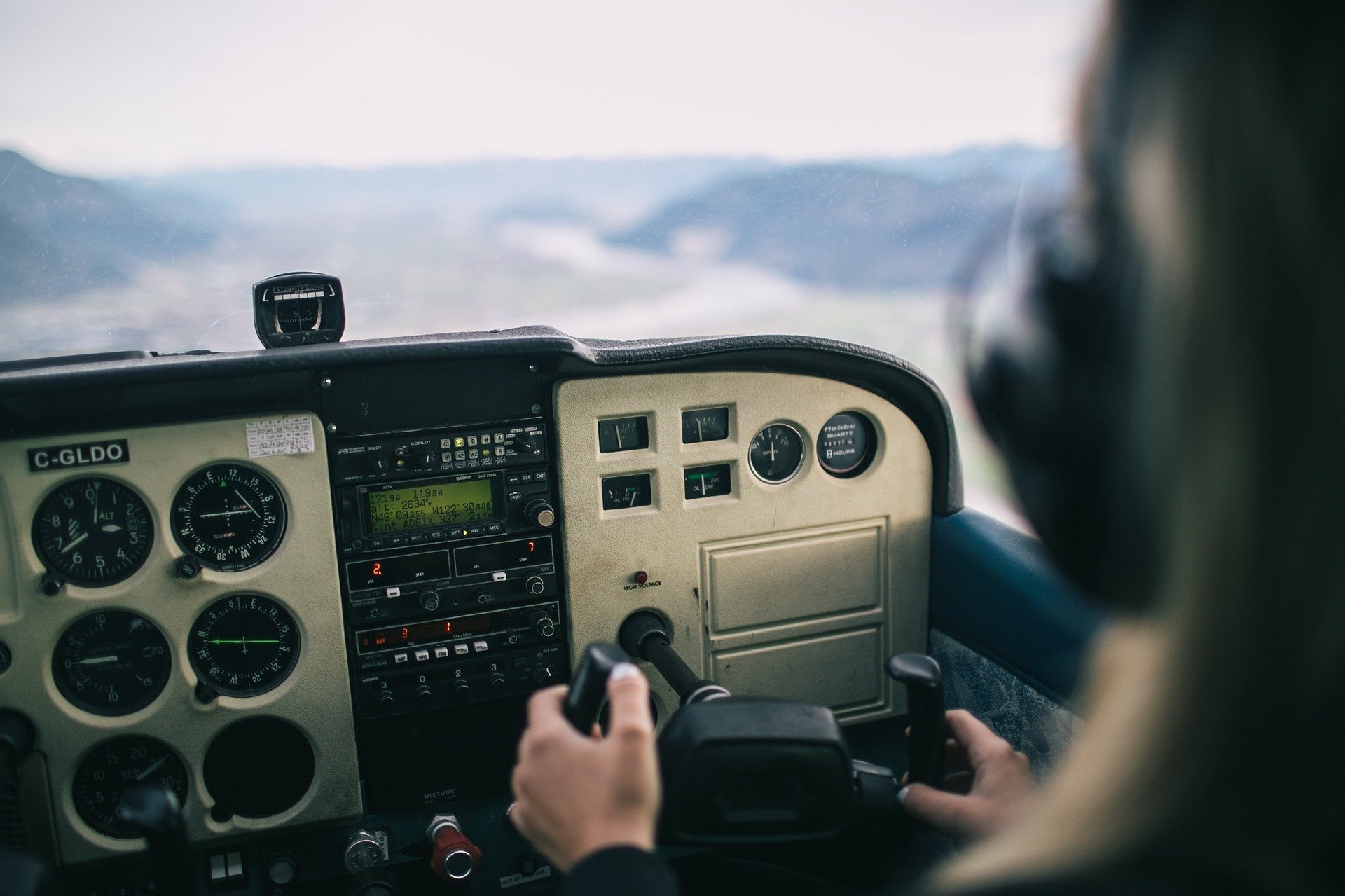}
\caption{Helicopter cockpit. The high number of visual displays and the inevitable use of headphones pave the way for psychoacoustic sonification for navigation.}
\label{pic:helicopter}
\end{figure}

Aircraft tend to have a very high number of visual displays. This is true for civil and military airplanes, jets and helicopters, as illustrated in Figs. \ref{pic:civil} to \ref{pic:helicopter}.

These displays serve to fly the aircraft, referred to as \emph{instrument flying}. Instruments include altitude meter, airspeed indicator, vertical speed indicator, turn and slip indicator, compass or another heading indicator, and attitude indicator, indicating pitch and roll. Instrument flying is particularly necessary in bad visual conditions, e.g., in darkness, clouds, fog, smoke or heavy rains. Unfortunately, instrument flying imposes a ``tremendous burden'' on the eyes \cite{blindflight} and is cognitively demanding, since the information of multiple displays needs to be interpreted and integrated. During some maneuvers the pilot's ability to interpret visual displays or match visual and balance perception may be impaired \cite{airplane,blindflight,jmui}. Consequently, spatial disorientation causes most aviation accidents \cite{airplane}.

Auditory displays have been developed to enable blind flight \cite{blindflight,airc,airplane}, i.e., stable flight without visual cues. These studies conclude that it would be highly desirable to have a system that communicated pitch and roll with a high resolution on a linear ratio scale of measure, i.e., with an obvious coordinate origin that represents the horizon. Psychoacoustic sonification fills exactly this gap. Here, the coordinate origin represents the horizon, two dimensions represent pitch and roll of the attitude indicator, which is the primary instrument for flying. The third dimension may represent the air speed. Additional earcons can serve to represent altitude. This even enables pilots to stay on a pre-defined course or carry out maneuvers safely, without the need for visual feedback.

\subsection{Autonomous Vehicles}
\begin{figure}[thbp]
\centering
\includegraphics[width=0.45\textwidth]{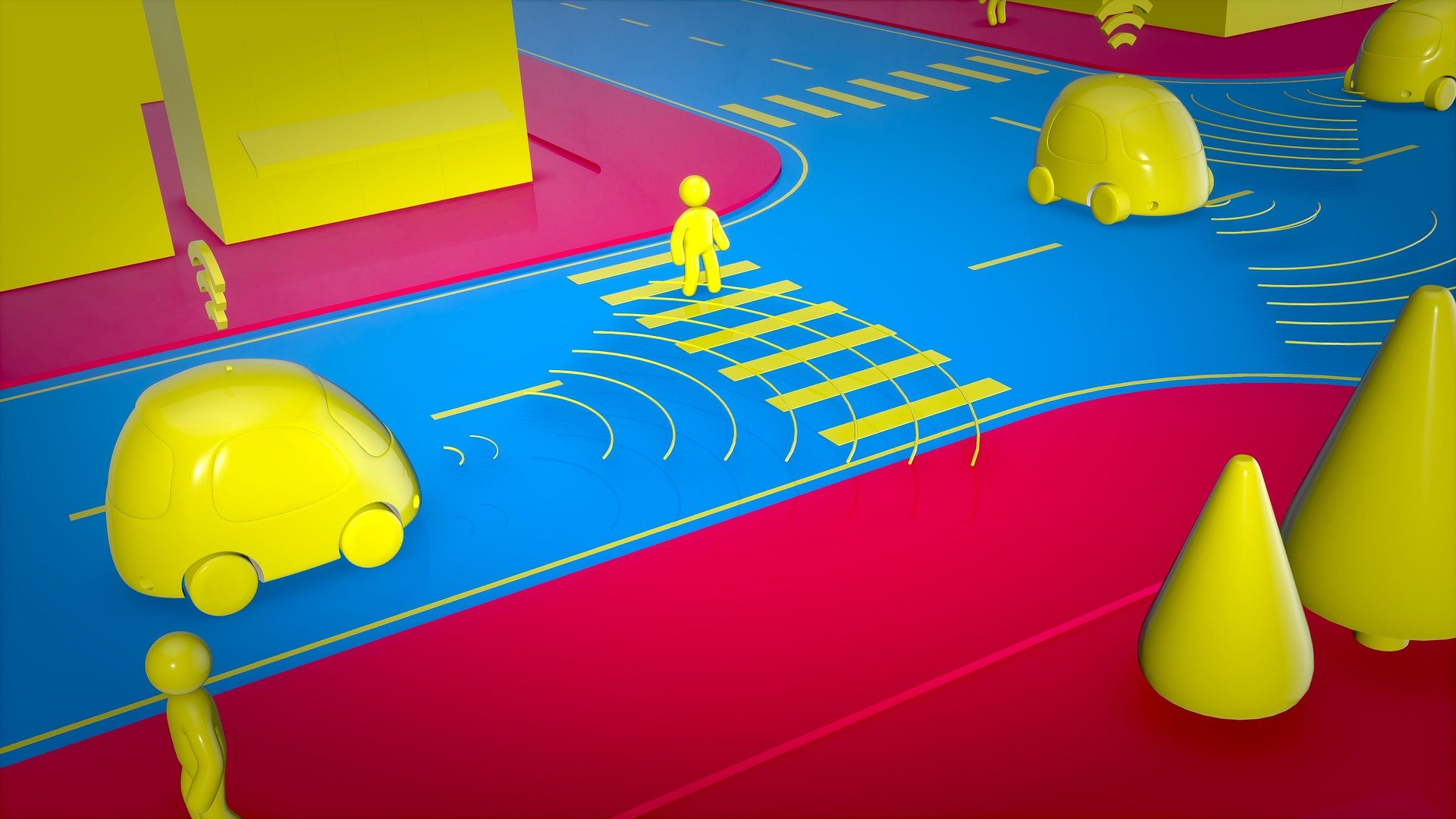}
\caption{Autonomous vehicles of level 0 to 3 require the driver to stay aware of the car action and readily intervene when called upon by the vehicle.}
\label{pic:auto}
\end{figure}

Autonomous vehicles, like self-driving cars, are classified into six different levels of autonomy by the Society of Automotive Engineers (SAE) \cite{sae}. Only levels 4 and above allow the driver to be inattentive. At lower levels the environment may be scanned by means of sensors as illustrated in Fig. \ref{pic:auto}. The driver needs to stay aware of the vehicle's actions to take over control if necessary.

Here, hand-over and take-over commands and alarms are the minimum requirement. However, psychoacoustic sonification can inform the driver continuously about the actions of the vehicle. For example, the automotive can spontaneously decide to deviate from the planned route if it receives information on a traffic jam or recognized a ``Road closed!'' sign. Here, one sonification dimension could indicate the pending deviation, i.e., that the car plans to turn right instead of driving straight ahead. Another sonification dimension could indicate the absolute deviation from the planned route. The presence of the sound capzures attention and increases driver awareness. Furthermore, the driver is informed about the decision of the vehicle and its consequence for the route. These pieces of information help the driver decide whether it is acceptable and necessary to turn right, and if the new route seems appropriate or causes too much deviation. If preferred, the driver can take over control. Here, one characteristic of the above-described psychoacoustic sonification becomes effective: the larger the deviation from the planned path and route, the more obtrusive and unpleasant the sound becomes for the listener. This makes the sonification intuitive on a qualitative level.

In a similar fashion, the psychoacoustic sonification could inform the driver about the confidence level of the implemented artificial intelligence. For example, if road signs or road markings are recognized with a low confidence, the sound could pass this information to the driver. Then, the driver is aware about the general confidence and confidence change but can still concentrate on other things until he or she decides to take action. This is the advantage of psychoacoustic sonification over speech messages, which require more attention and more cognitive resources to be interpreted.

%obstacles and targets may be tracked by means of Very-High Frequency Omnidirectional Range or Nondirectional Radio Beacon coupled to a gyrocompass and equipped with two needles to indicate position of target

%6 basics:
%airspeed indicator, attitude indicator, altimeter, turn coordinator, heading indicator, and vertical speed indicator
\subsection{Image-Guided Surgery}
\label{surgery}
Image-Guided Surgery (IGS) enables surgeons and other clinicians to carry out interventions that are too risky without guidance. Examples are biopsy of tissues or resections of tumors near critical structures, like large arteries, important nerves, or sensitive membranes. %A special type of IGS is Neuronavigation, where small 

In image-guided surgery (IGS), the human or veterinary patient is scanned, e.g., by means of computed tomography or magnetic resonance imaging. From the scan, an augmented reality 3D model is created. It shows the anatomy, where associated structures are grouped, like vessels, nerves, bones, healthy tissue and tumor. Each group can receive an individual color and opacity level. This helps to localize the exact location of the tumor, the center and the extent. With this model clinicians plan the surgery, e.g., where to make incisions and how to approach the tumor past critical or impenetrable structures. These plans, may include insertion points, and the angles and depth of an ablation needle from the surface to the center of the tumor. Or they include the cutting trajectory for a scalpel blade.

During the procedure, the location and orientation of the surgical tool is tracked in relation to the patient anatomy. Tracking devices are electromagnetic sensors or optical markers that are sensed by stereoscopic infrared cameras. Tools include scalpels, cauterizers, biopsy or ablation needles. On a screen clinicians see an overlay of the surgical tool and the augmented anatomy. This mixed reality situation guides the clinician, who can see the current location of the tool, the target location, and the position of obstacles, i.e., critical and impenetrable structures. In terms of HMI, the tool is the HID, the tracking system is the sensor and the screens show the UI.

\begin{figure}[thbp]
\centering
\includegraphics[width=0.45\textwidth]{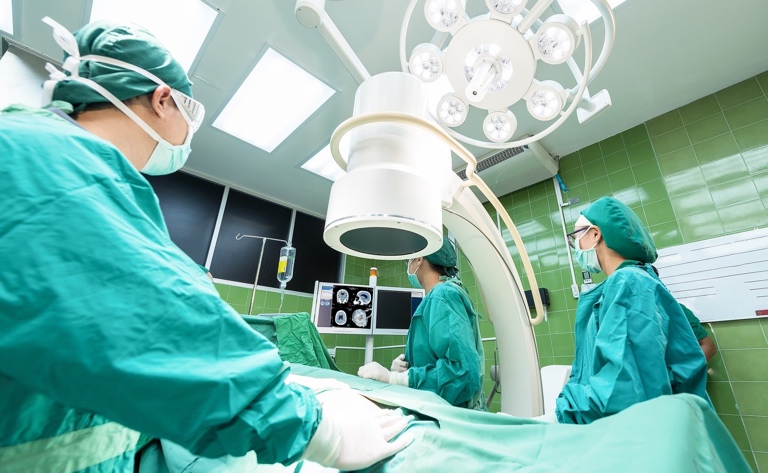}
\caption{Typical IGS situation: The monitor is too far away to see details. It shows an overlay of the surgical tool in a pseudo-3D graphic, and three 2D graphics that show the three anatomic planes. None of the graphics represents the first-person perspective of the surgeon of the assistants. Still, they have to take their eyes off the patient to navigate during the surgery.}
\label{pic:igs}
\end{figure}

% add BMC citation somewhere here
The major drawback of IGS is the high cognitive demand that mostly originates in the UI design. Typical IGS setups (see  Fig.~\ref{pic:igs} for an example) give rise to a number of spatial challenges such that mentally processing spatial information becomes a crucial ability for successfully performing IGS \cite{vajsbah.schulth:spati}. Often, installing the monitors has a low priority, so they might be located a little far away or even behind the surgeon, which is rather unergonomic. The screen shows a pseudo 3D model of the patient together with the three anatomic planes of the patient, i.e., sagittal, coronal and axial. All four graphics are overlain by a depiction of the surgical tool. However, none of the graphics' perspectives coincides with the perspective of the surgeon. So he or she has to mentally integrate the four graphics, mentally rotate, translate and scale the them to navigate properly. This is quite demanding. For example, the surgeon might move the tool to her right. This might cause a motion of the overlain tool to the upper left in the pseudo 3D-model, a small motion to the upper right in the median view, a motion far to the right and slightly up in the coronal view and a motion to the right and slightly down in the axial view.

In addition to that, 3D graphical processing can cause latencies of hundreds of milliseconds. This forces clinicians to operate so slowly that they can always anticipate their proximity to target and obstacles, despite the delayed feedback.

Here, psychoacoustic sonification offers a huge advantage. Instead of looking at a screen to derive spatial information from integrating four seemingly conflicting views, the surgeon can visually focus on the patient while navigating by means of sound. The sound can either communicate the location of the tool in relation to the planned trajectory, location of the nearest obstacle, or the location of the target in relation to the current tool position.

\subsection{Anesthesiology}
In the medical sector, not only surgeons can benefit from psychoacoustic sonification. Anesthesiologist monitor vital parameters of patients in the operating room, the postanesthesia care unit and the intensive care unit. Vital parameters include heart rate, blood pressure, end-tidal carbon dioxide (capnography), respiration rate, tidal volume, oxygen concentration in the blood, and body temperature \cite{eyesfree}. These tend to have an ideal value, a healthy range, and critical thresholds that depend on factors, like age, type of surgery, disease state, and anesthetics \cite{ana}.

\begin{figure}[thbp]
\centering
\includegraphics[width=0.45\textwidth]{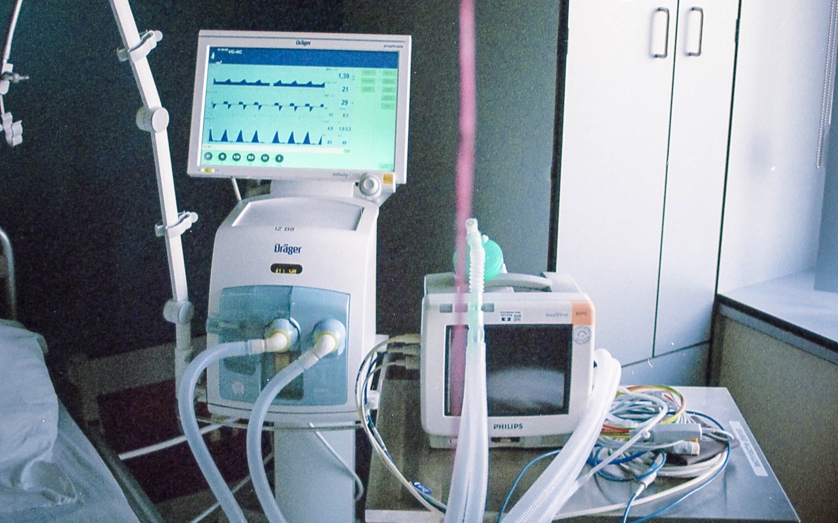}
\caption{User interface for monitoring a patient's vital functions. A monitor is combined with an auditory display to serve as a multi-modal user interface.}
\label{pic:ana}
\end{figure}

A typical user interface for vital parameter monitoring is illustrated in Fig. \ref{pic:ana}. The current state and a short history of multiple vital functions are plotted over time and displayed on a screen. Often, integrated loudspeakers play additional pulse-oximetry sounds. %To record the data, they plots may be printed on a paper roll.

Auditory pulse oximeters are among the most widely used sonifications. They inform the anesthesiologist about the heart rate and the relative oxygen concentration in the blood. The strength of this auditory adjunct to the visual UI is that it enables anesthesiologists to keep monitoring and staying aware of two essential parameters without looking at a screen, e.g., while counteracting adverse reactions, intubating, or repositioning the patient. There are attempts to add information of even more vital functions \cite{heart,eyesfree}. The main drawback of conventional sonification for patient monitoring is that the sounds fail to communicate precise quantitative information about the relation between current state, target value and/or critical thresholds.

In terms of HMI, sensors measure the vital functions over time. Visual displays, i.e., plots on a screen, serve as UI, together with pulse-oximetry sounds. Psychoacoustic sonification can serve to communicate the state of one or more vital functions in relation to the target value and/or critical thresholds. This has been demonstrated for a pulse-oximetry scenario of neonates \cite{schwarzziemer}. Here, users are able to identify whether the oxygen saturation lies safely within the target region (near $92.5$\%, near or within the critical region below $90$\% or above $95$\%). % In contrast to the scenarios discussed above, patient's vital functions are not multidimensional, but multivariate.
%Means to increase body temperature range from heating and humidification of anesthetic gases over passive insulation or active circulating-water mattresses, to warming intravenous fluids.

\subsection{Water Quality Monitoring}
In regions with chlorinated water distribution systems, households and pubic drinking fountains can be equipped with electrode-based pH meters, digital chlorine sensors and thermometers to measure pH, residual chlorine and temperature of the tap water in real-time. Such sensors serve as a fallback solution in addition to the monitoring that is required for distribution systems of municipal water systems.

\begin{figure}[thbp]
\centering
\includegraphics[width=0.45\textwidth]{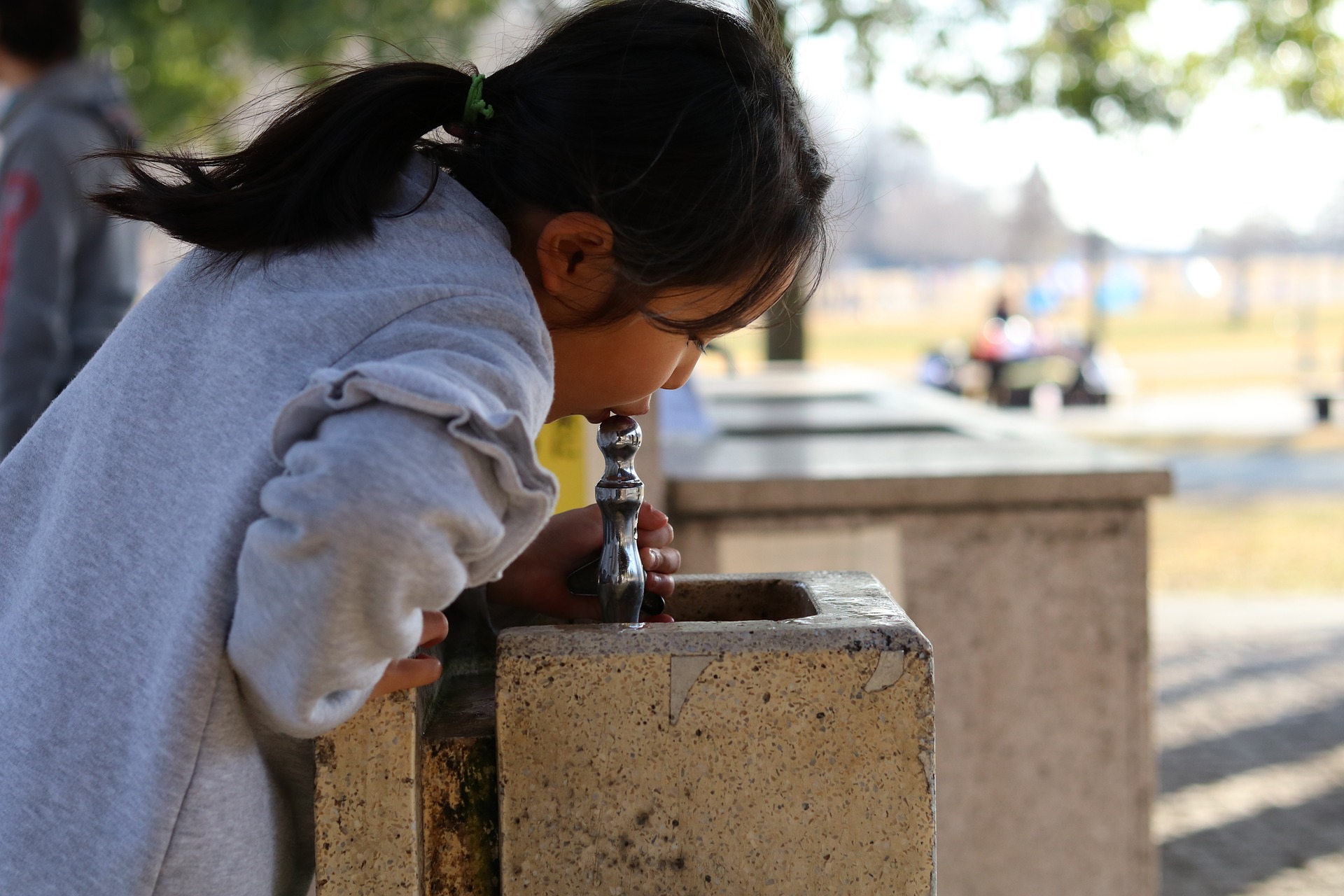}
\caption{While drinking water, an obtrusive and unpleasant sound will obvious .}
\label{pic:water}
\end{figure}

Raising instant awareness of these measures can help to prevent from water-borne diseases and protect children from tap water scald burns. However, as illustrated in Fig. \ref{pic:water}, the typical drinking posture prevents you from looking at a screen. Here, psychoacoustic sonification is the ideal UI. As described above, it can be equipped with an innate quality, where large deviations from the coordinate origin sound unpleasant. Here, a clean sound represents optimal drinking water conditions at the coordinate origin. Deviations sound sound either rough or siren-like; both intuitively sound alarming.

Water quality monitoring is just one out of a number of monitoring tasks that can be assisted by psychoacoustic sonification. We can also imagine to monitor temperature and flow rate of a cooling circuit, visits, downloads and responds on a web-server or pressure, temperature and wind velocity of a weather station.
\subsection{Firefighters}
\begin{figure}[thbp]
\centering
\includegraphics[width=0.45\textwidth]{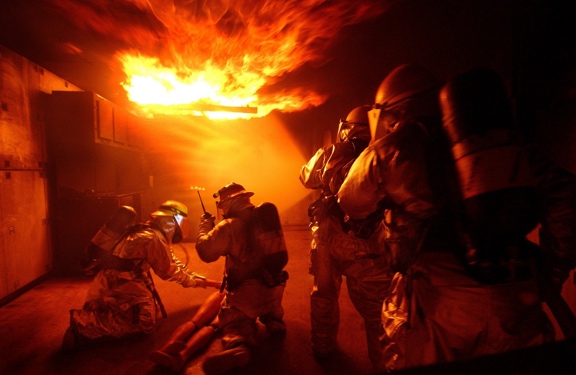}
\caption{On rescue missions inside burning buildings, firefighters use sensors to recognize obstacles in their environment or holes in the ground. Psychoacoustic sonification can warn them about the nearest danger to guide them safely.}
\label{pic:fire}
\end{figure}

On a rescue mission, firefighters may enter burning or smoke-filled building. An exemplary training scenario can be seen in  Fig. \ref{pic:fire}. Here, firefighters can hardly see their environment. Approaches exist to equip firefighters with sonar \cite{rodents} or radar \cite{radfir} sensors. These localize obstacles or holes in the ground in smoke-filled buildings. Due to the limited vision, even hand-held monitors cannot be employed as a UI to pass the sensor information to the firefighter. Therefore, haptic devices \cite{rodents} \cite{radfir} have been suggested. These inform the firefighter about the presence of an obstacle around or a hole in the ground to increases their spatial awareness.

In this scenario, psychoacoustic sonification can not only inform them about the presence of an obstacle or hole, but even about its location, i.e., direction and distance. This additional information allows them to explore the building faster and to distribute more widely compared to haptic 'presence only'-feedback and compared to conventional exploration along a guide-rope.

Note that firefighters often use drones to get an overview, to monitor and coordinate operations in disaster areas, like flood, wildfire and large-scale structure fires, and to search and rescue victims. Likewise, they may fly firefighting planes or  helicopters. One sound, i.e., one type of psychoacoustic sonification could serve as a UI for all these application areas of firefighters.

\subsection{Freight Container Cranes}
\begin{figure}[thbp]
\centering
\includegraphics[width=0.45\textwidth]{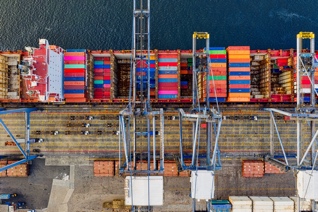}
\caption{Freight container cranes in a container port. The large distance between operator and container as well as the high density of piles make visual navigation demanding and dangerous. Psychoacoustic sonification can inform crane operators about the distance and direction of the nearest container to avoid collisions.}
\label{pic:cranes}
\end{figure}

In container ports, freight container cranes unload containers from container ships. These containers are densely piled. Due to continual growth of container ships, the containers need to be piled faster and more densely every year. A photo of a container port is shown in Fig. \ref{pic:cranes}.

Piling cranes based on vision is very demanding. Due to the large distance between the crane operator and the container, it may be difficult to estimate whether spaces are sufficiently large for a container to get through. Especially depth perception and estimation of spatial constellations in partly occluded scenes is demanding. Furthermore, the steel cabin as well as other containers and cranes may party occlude the visual field. To assist crane operators in piling the containers without accidental collisions, cranes became equipped with proximity sensors, such as laser rangefinders. If an obstacle is dangerously close, an alarm will sound. In this situation psychoacoustic sonification could offer much more than just an alarm. The sound could inform the operator not only about the fact that an obstacle is close. It even informs the operator where, i.e., in which direction and how far, the obstacle lies.

%aircrafts and vessels, like helicopters, civil airplanes and fighter aircrafts, spacecrafts, missiles, submarines, underwater vehicles, welding robots, automotives, remote vehicle control of unmanned ground freeride vehicle, image-guided surgery and neuronavigation in human and veterinary medicine.

%multivariate data analysis

%New Interfaces for Musical Expression (NIME) 

%whenever the location of a a target or an obstacle is defined --- e.g., route planing, image-guided surgery and neuronavigation, in Virtual Reality(VR) applications, games or flight simulators --- or detected, e.g., by means of sensors.

\section{CONCLUSION}
In this paper we gave an extensive introduction to \emph{psychoacoustic sonification}. It is allocated at the user interface part of the human-machine interaction circle. The psychoacoustic closed loop can serve as auditory user interface or as the acoustic part of a multimodal user interface. Psychoacoustic sonification takes the largely interfering and nonlinear relationships between acoustics and auditory perception into account in the digital signal processing for the sound design. This ensures interpretability, linearity, continuity, a high resolution and absolute magnitudes of each single dimension in an uni-or multidimensional sonification. Furthermore, psychoacoustic sonification ensures orthogonality and integratability of the axes. This means, psychoacoustic sonification can carry more information than conventional sonification, without losing any clarity.

We referred to our previous publications on the theory, implementation and experimental evaluation of psychoacoustic sonification and reported the state of the art and ongoing research. Moreover, we explained what benefits psychoacoustic sonification can offer for human-machine interaction.

Psychoacoustic sonification is beneficial whenever vision is the bottleneck of performance, be it due to limited vision, visual overload, demanding visual interpretation or the limited focal point of visual attention. This is the case in various scenarios in all kinds or application areas.

Researchers and practitioner of many disciplines are unaware of sonification and the benefit that auditory displays can offer to them. Hence, we presented a number of practical use cases from multiple application areas where the psychoacoustic sonification can take full effect. Of course, our listing is not exclusive, but exemplary.

\section{OUTLOOK}
We are still in the progress of evaluating the benefit of the psychoacoustic sonification as a user interface for human-machine interaction in both abstract scenarios and a practical use case, namely surgical navigation. Pretests by the authors, demo sessions on conference and the initial results of the experiments provide evidence that the 3D psychoacoustic sonification enables users to navigate through three-dimensional space after just a couple of minutes of training. Note that the precise sonification of three orthogonal and continuous dimensions is not a natural upper limit. It is possible to include three new dimensions not based on a tonal but a noisy sonification signal. Together, such a six-dimensional sonification could allow an operator to navigate to a target location in three-dimensional space while avoiding obstacles. Or the six dimensions could represent all six degrees of freedom of an object, including three-dimensional location and orientation. For blind flight, new dimensions could sonifiy the information of the turn and slip indicator, enabling pilots to carry out a number of complicated maneuvers blindly.

%We already have ideas to .

Note that the psychoacoustic sonification might be optimized for each individual application. Ideally, as little information as needed is provided aurally, to shorten the learning time. 

% \cite{bruce}.
\bibliographystyle{plainurl}
\setcitestyle{square,numbers}
%\bibliographystyle{apa}
%\bibliographystyle{plainurl}%ieeetr,IEEEtran
%aipnum4-1 does not provide paper titles 
%apa is -alphabetic, -no DOI, +abbreviates first names.
%bestpapers only 5 references show up
%ieeetr is +unsorted and +abbreviates first names, but -lacks DOI
%IEEEtran is +unsorted and +abbreviated first names, but only -delivers URL, not DOI
%myIEEEtran does not link DOI, ---
%plainurl is alphabetic, but has DOI
%plainnat is alphabetic, but has DOI
%pnas does not provide title
%ppcf no doi
%unsrt is unsorted but lacks DOI
\bibliography{iwin}

\flushleft{
(Received October XX, 2008)\\
(Revised July XX, 2009)
}

\INFSOCbiography[ziemer]{Tim Ziemer}{%
is a researcher in the field of applied psychoacoustics. He received his doctorate in Systematic Musicology at the University of Hamburg in 2015. Since then, he worked as a researcher at the Institute of Systematic Musicology at the University of Hamburg, at the National Institute of Informatics, Tokyo, and in the Medical Image Computing Group and the Bremen Spatial Cognition Center at the University of Bremen. His research interests include sound field synthesis, sonification, microphone array technology and music information retrieval. He received several scholarships and awards, e.g., by the Acoustical Society of America (ASA), Claussen Simon Foundation, Association for Computing and Machinery (ACM) and Special Interest Group on Information Retrieval (SIGIR), and by Hyundai Motors and the International Community for Auditory Display (ICAD).}

\INFSOCbiography[author2]{Nuttawut Nuchprayoon}{%
is a PhD candidate and equine practitioner at the Faculty of Veterinary Science, Mahidol University, in Nakhon Pathom, Thailand. He has started working as equine clinician in 2014. His research interests are equine biomechanics and equine dentistry. He got an intern abroad scholarship for short-term studying in Germany during June to August 2019 at Bremen Spatial Cognition Center, University of Bremen.}

\INFSOCbiography[a5]{Holger Schultheis}{has a background in Computer Science and Cognitive Psychology and is currently a senior researcher at the University of Bremen. His research aims to bridge the gap between natural and artificial intelligence. By combining insights from cognitive science and methods from artificial intelligence he works towards more sophisticated artificial agents and a deeper understanding of human cognition. Among other things, he applies the insights gained from his basic research to build intelligent tutoring and cognitive assistance systems that flexibly adapt to the current needs of the human user.
}

\end{document}